\documentclass[aip,jap,twocolumn,reprint]{revtex4-1}
\pdfoutput=1
\usepackage[squaren, thinqspace]{SIunits}
\usepackage{amsmath}
\usepackage{amssymb}
\usepackage{graphicx}
\usepackage{xcolor}
\usepackage{comment}
\usepackage[caption=false]{subfig}
\usepackage[colorlinks=true, linkcolor=blue, citecolor=blue, filecolor=blue, pagecolor=blue, urlcolor=blue]{hyperref}

\newcommand{\fermiint}[2]{\mathcal{F}_{#1}(#2)}
\newcommand{\fermiquot}[3]{H_{#2}^{#1}(#3)}

\begin{document}

	\title{Laser damage in silicon: energy absorption, relaxation and transport}
	\author{A. \surname{R\"{a}mer}}
	\email{araemer@physik.uni-kl.de}
	\affiliation{Department of Physics and Research Center OPTIMAS, University of Kaiserslautern, 67663 Kaiserslautern, Germany}
	\author{O. \surname{Osmani}}
	\affiliation{Faculty of Physics, University of Duisburg-Essen and CeNIDE, 47048 Duisburg, Germany}
	\author{B. \surname{Rethfeld}}
	\affiliation{Department of Physics and Research Center OPTIMAS, University of Kaiserslautern, 67663 Kaiserslautern, Germany}
	\date{\today}

	\begin{abstract}
		Silicon irradiated with an ultrashort \unit{800}{\nano\meter}-laser pulse is studied theoretically
		using a two temperature description that
		considers the transient free carrier density during and after irradiation. A Drude model is implemented to account for the
		highly transient optical parameters. We analyze the importance of considering these density-dependent parameters
		as well as the choice of the Drude collision frequency.
		In addition, degeneracy and transport effects are
		investigated. The importance of each of these processes for resulting calculated damage thresholds is studied.
		We report damage thresholds calculations that are in very good agreement with experimental results over 
		a wide range of pulse durations.
	\end{abstract}
	\maketitle

	\section{Introduction}
	The interaction of laser pulses with solid matter has been subject of both
	experimental\cite{Preston84,Wang94,Goldman94,Chichkov96,Sjodin98,S-T00,Sabbah02,Allenspacher03,Englert08} 
	and theoretical\cite{Anisimov74,vanDriel87,Rethfeld99,Rethfeld04,Bulgakova04,Bulgakova05,Christensen09,Autrique13} 
	research for many years. Especially ultrashort laser pulses, that cause minimal collateral damage to the surrounding 
	material, are of high interest for applications in medical surgery, micromachining and nanostructuring.\cite{Vogel03,Li03,Rodriguez09,Baeuerle}
	
	For the theoretical description of laser-excited solids, both the timescale and the kind of material have to be
	carefully considered.\cite{Rethfeld04APA} During and directly after the excitation, the electron system is out of equilibrium and
	no temperature is defined. Consequently, nonequilibrium descriptions like the Boltzmann
	equation\cite{Bejan97,Rethfeld99,Kaiser00,Pietanza04,Pietanza07,Mueller13,Shcheblanov13} 
	or kinetic Monte Carlo simulations\cite{Gao07NIMB, Gao07NIMA, Medvedev11, Laporta11} have to be applied.

	On a timescale of about a hundred femto\-seconds\cite{vanDriel87,Goldman94,Mueller13} 
	after excitation, the electrons again follow an equilibrium distribution. A temperature can be assigned 
	to both the electrons and the lattice. However, those temperatures will differ 
	because the laser energy is mainly absorbed by the electrons and subsequently transferred to the lattice.
	Equilibration between both subsystems happens on a picosecond timescale. For metals, the well known two temperature
	model\cite{Anisimov74} is often applied on this timescale to describe both heat relaxation and transport within the
	electron and lattice subsystem.

	In contrast to the excitation of metals, it is not sufficient to follow carrier and	lattice temperatures
	when describing laser-excited semiconductors. Because the free carrier density in these materials can vary 
	over several orders of magnitude during irradiation, its evolution has to be treated explicitly. 
	In addition, the highly transient free carrier density causes huge changes in the optical properties
	of the solid. Thus, a theoretical description of laser excitation of semiconductors has to account for the transient free carrier density,
	transient optical properties, carrier and lattice temperature evolution as well as energy and particle transport.

	This paper extends an existing two temperature description for semiconductors first presented by van Driel\cite{vanDriel87} 
	to account for the transient optical properties during the excitation with a femtosecond-laser pulse. 
	To that end, we will first introduce the theoretical model and the extension. Afterwards, we will investigate 
	the influence of the transient optical parameters as well as degeneracy and transport effects on the results of 
	the simulations. We will discuss the importance of each property and process under investigation by calculating 
	damage thresholds and comparing with experimental data.

	\section{The Density-Dependent Two Temperature Model\label{sec:nTTM}}
		In laser-irradiated semiconductors,	electrons are excited from the valence to the conduction band via single or	multiphoton 
		absorption. Thus, electron-hole pairs are created. The order of the absorption process depends on the photon
		energy in relation to the band gap energy of the material. In our case, the irradiation of silicon with
		\unit{800}{\nano\meter}-laser pulses, the photon energy is larger than the indirect band gap of silicon. 
		Here, it is sufficient to consider single and two photon absorption processes.
		Higher order processes are less probable and can, thus, be neglected.\cite{Bristow07}
		In addition, free carriers 
		(electrons in the conduction band and holes in the valence band) can absorb further photons via free carrier absorption
		thereby increasing their kinetic energy. Conduction band electrons with a sufficiently high kinetic energy may
		ionize additional valence band electrons via impact ionization. Furthermore, free electrons and holes may recombine via Auger
		recombination transferring the excess energy to another free electron or hole. Meanwhile, the carrier system can
		couple to the lattice system. The temperatures of both systems will equilibrate until a thermal equilibrium state is
		reached.

		All these processes as well as carrier and heat diffusion can be modeled with the density-dependent two temperature model
		(nTTM) first presented by van Driel\cite{vanDriel87}. For sake of completeness, we will in the following
		derive the original model following his work\cite{vanDriel87} 
		before expanding it to account for transient optical parameters during irradiation.
		
		\subsection{Basic Assumptions}
			In the framework of the nTTM, free electrons and holes are assumed to be thermalized into an equilibrium distribution
			function, usually a Fermi distribution
			\begin{align}
				f_c(\varepsilon) = \frac{1}{1+\exp\left[\frac{\pm(\varepsilon-\mu_c)}{k_BT_e}\right]}\,,
				\label{eq:Fermi_distribution}
			\end{align}
			at all times. The distribution functions of electrons and holes are assumed to posses different chemical potentials 
			$\mu_c$, where $c$ stands for ``carrier'' and may be substituted by $e$ for electrons and $h$ for holes, but a common 
			temperature $T_e$. 	Here, $k_B$ denotes the Boltzmann constant.	The plus sign in the exponent in Eq.~\eqref{eq:Fermi_distribution} 
			associates with the electrons, while the minus sign	associates with the holes. 
			The assumption of thermalized carriers is, however, questionable at times shortly after
			laser-excitation.
			As thermalization will take on the order of \unit{100}{\femto\second},\cite{vanDriel87,Goldman94,Mueller13}
			carrier temperature and distribution functions as well as chemical potentials may be interpreted as quasi-temperature, 
			quasi-Fermi distribution functions and quasi-chemical potentials.
			
			Conduction and valence band	are treated as parabolic bands. Consequently, the densities of states (DOS) of electrons 
			\begin{subequations}
				\label{eq:DOS}
			\begin{align}
				&\quad				&D_e(\varepsilon)	&= \frac{m_{e,\mathrm{DOS}}^{*3/2}}{\hbar^3\pi^2}\,\sqrt{2\,(\varepsilon-\varepsilon_{C})}\\
				&\text{and holes}	&D_h(\varepsilon)   &= \frac{m_{h,\mathrm{DOS}}^{*3/2}}{\hbar^3\pi^2}\,\sqrt{2\,(\varepsilon_V-\varepsilon)}
			\end{align}
			\end{subequations}
			are applied, where $\varepsilon_C$ and $\varepsilon_V$ denote the lower edge of the conduction
			band and the upper edge of the valence band, respectively. A different constant	DOS
			effective mass $m_{c,\mathrm{DOS}}^{*}$ is assumed for electrons and holes, respectively, see Tab.~\ref{tab:model_parameters}.

\begin{figure}
				\includegraphics{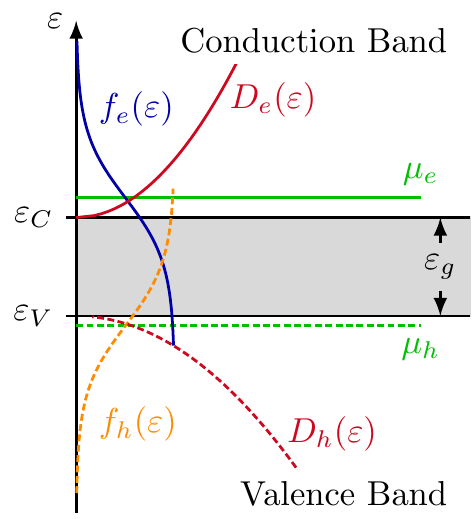}
				\caption{
				(Color online) Sketch of DOS, quasi-Fermi distribution and quasi-chemical potentials of electrons and holes.
				\label{fig:bandscheme}}
			\end{figure}

			Figure~\ref{fig:bandscheme} shows a sketch of the DOS of electrons and holes as well as their quasi-Fermi
			distributions and chemical potentials.

			By integrating over the carrier distribution multiplied with the respective DOS, the local carrier density reads
			\begin{align}
				n_c = 2\,\left(\frac{m_{c,\mathrm{DOS}}^*\,k_BT_e}{2\pi\,\hbar^2}\right)^{3/2}\,\fermiint{1/2}{\eta_c}\,.
				\label{eq:local_density}
			\end{align}
			Here, the function
			\begin{align}
				\fermiint{\xi}{\eta_c} = \frac{1}{\Gamma(\xi+1)}\,\int_0^\infty\frac{x^\xi}{1+\exp\left(x-\eta_c\right)}\,\mathrm{d}x
				\label{eq:fermiint}
			\end{align}
			denotes the Fermi integral of order $\xi$. The so-called reduced Fermi levels of electrons and holes are defined as
			\begin{align}
				\eta_e = \frac{\mu_e-\varepsilon_C}{k_BT_e} \text{\qquad and\qquad} \eta_h=\frac{\varepsilon_V-\mu_h}{k_BT_e}\,,
				\label{eq:reduced_fermi_level}
			\end{align}
			respectively.\cite{vanDriel87} They indicate the position of the respective chemical potentials in
			relation to the band edges. If the reduced Fermi levels	become positive, the respective chemical potentials are
			positioned inside the respective bands and the carrier system is degenerate.
		
		\subsection{Carrier and Energy Current}
			In a laser-generated free carrier system, electrons and holes basically move together as electron-hole pairs since
			the Dember field that builds up	due to charge separation prohibits carrier and current densities from becoming significantly 
			different (ambipolar diffusion).\cite{vanDriel87}
			We can therefore assume, that 
			\begin{align}
				\label{eq:ambipolar_diffusion}
				n_e=n_h \text{\qquad and\qquad} \vec{j}_e=-\vec{j}_h
			\end{align}
			at each point in space. Under these conditions, the carrier current density is given by
			\begin{align}
				\vec{J} &= -D\Bigg\{\vec{\nabla}n_e 
							+ \frac{n_e}{k_BT_e}\left[\fermiquot{1/2}{-1/2}{\eta_e}+\fermiquot{1/2}{-1/2}{\eta_h}\right]^{-1}\vec{\nabla}\varepsilon_g\nonumber\\
						&\quad	+ \frac{n_e}{T_e}\left[2\,\frac{\fermiquot{1}{0}{\eta_e}+\fermiquot{1}{0}{\eta_h}}{\fermiquot{1/2}{-1/2}{\eta_e}+\fermiquot{1/2}{-1/2}{\eta_h}}-\frac{3}{2}\right]\,\vec{\nabla}T_e
							\Bigg\}\,,
				\label{eq:carrier_pair_current_final}
			\end{align}
			where $\varepsilon_g=\varepsilon_C-\varepsilon_V$ denotes the band gap 
			energy, $\fermiquot{\xi}{\zeta}{\eta_c}=\fermiint{\xi}{\eta_c}/\fermiint{\zeta}{\eta_c}$ is the ratio of Fermi
			integrals~\eqref{eq:fermiint}
			and 
			\begin{align}
				D &=	\frac{k_BT_e}{e}\,
						\frac{\mu_e^0\mu_h^0\,\fermiquot{0}{1/2}{\eta_e}\,\fermiquot{0}{1/2}{\eta_h}}{\mu_e^0\,\fermiquot{0}{1/2}{\eta_e}+\mu_h^0\,\fermiquot{0}{1/2}{\eta_h}}\nonumber\\
				  &\quad\times\left[\fermiquot{1/2}{-1/2}{\eta_e}+\fermiquot{1/2}{-1/2}{\eta_h}\right]
				\label{eq:ambipolar_diffusion_coefficient}
			\end{align}
			with the carrier mobility $\mu_c^0$
			is the ambipolar diffusion coefficient. Note that the band gap gradient	$\vec{\nabla}\varepsilon_g$
			is non-zero because of band gap shrinkage due to
			increased lattice
			temperature and carrier density during and after excitation.\cite{Varshni67,S-T00,Jellison83}

			The heat current density $\vec{W}=\vec{w}_e+\vec{w_h}$ is the sum of electron and hole
			heat current densities and reads
			\begin{align}
				\vec{W} &=	\left\{\varepsilon_g+2k_BT_e\,\left[\fermiquot{1}{0}{\eta_e}+\fermiquot{1}{0}{\eta_h}\right]\right\}\,\vec{J}
							\nonumber\\
						&\quad	- (\kappa_e+\kappa_h)\,\vec{\nabla}T_e\,.
				\label{eq:ambipolar_energy_current}
			\end{align}
			
			The expressions for the current densities include treatment of the Peltier and Seebeck effect as well as Ohm's
			and Fourier's law. A more detailed description and background information on the transport terms is given by van
			Driel.\cite{vanDriel87}

			Note that whenever space charge effects cannot be neglected, for instance when studying Coulomb explosion, the
			assumption of ambipolar diffusion~\eqref{eq:ambipolar_diffusion} is not valid anymore. Transport in this case has
			been studied by Bulgakova~\emph{et al.}\cite{Bulgakova04, Bulgakova05}
		
		\subsection{Energy and Particle Balance}
			For the particle balance equation, interband laser absorption processes, namely single and two photon absorption,
			impact ionization, Auger recombination and carrier transport are considered:
			\begin{align}
				\frac{\partial n_e}{\partial t}	&= \frac{\alpha_\mathrm{SPA} I}{\hbar\omega_L} + \frac{\beta_\mathrm{TPA}\,I^2}{2\hbar\omega_L}
													+ \delta\,n_e - \gamma\,n_e^3 - \vec{\nabla}\cdot\vec{J}\,.
				\label{eq:particle_balance}
			\end{align}
			The first two terms describe single and two photon absorption, the third and fourth term stand for impact ionization 
			and Auger recombination and the last term describes carrier transport.
			Here, $I$ denotes the intensity of the laser pulse, $\hbar\omega_L$ the photon energy, $\alpha_\mathrm{SPA}$ and
			$\beta_\mathrm{TPA}$ the single and two photon absorption coefficients, respectively, $\delta$ the coefficient for impact
			ionization and $\gamma$ the one for Auger recombination.
			All coefficients used in this work are given in Tab.~\ref{tab:model_parameters}.

			The total carrier energy density is given by
			\begin{align}
				U_{e-h}	&= \int_{-\infty}^{\varepsilon_V} \varepsilon\,D_h(\varepsilon)\,f_h(\varepsilon)\,\mathrm{d}\varepsilon
							+\int_{\varepsilon_C}^\infty \varepsilon\,D_e(\varepsilon)\,f_e(\varepsilon)\,\mathrm{d}\varepsilon
							\nonumber\\
						&= n_e\,\varepsilon_g +	\frac{3}{2}\,n_e\,k_BT_e\,\left[\fermiquot{3/2}{1/2}{\eta_e}+\fermiquot{3/2}{1/2}{\eta_h}\right]\,.
				\label{eq:carrier_energy}
			\end{align}
			The first term in the second line represents the potential energy of the carrier pairs, while the second term
			gives the kinetic energy of electrons and holes. 
			
			The total carrier energy is modified by laser absorption, 
			transport and carrier-phonon coupling. Thus, the balance equation for carrier energy density is
			\begin{align}
				\frac{\partial U_{e-h}}{\partial t} &= (\alpha_\mathrm{SPA}+\alpha_\mathrm{FCA})\,I + \beta_\mathrm{TPA}\,I^2
				\nonumber\\
												&\quad	-\vec{\nabla}\cdot\vec{W} - g\,(T_e-T_{ph})\,,
				\label{eq:carrier_energy_balance}
			\end{align}
			where $\alpha_\mathrm{FCA}$ denotes the free carrier absorption coefficient. The carrier-phonon coupling parameter $g$
			is  non-constant in semiconductors and highly depends on carrier density.
			It is therefore often described by $g=C_{e-h}/\tau_\mathrm{c-ph}^\mathrm{relax}$, where the carrier heat capacity $C_{e-h}$ 
			depends on the carrier density and $\tau_\mathrm{c-ph}^\mathrm{relax}$ is the carrier-phonon energy-relaxation
			time.\cite{vanDriel87, Chen05, Gan12} 
			This expression results from an analytical solution of the traditional TTM for metal films at times after
			laser-excitation, neglecting transport and assuming a constant heat capacity as well as a constant electron-phonon coupling parameter.
			Obviously, it is therefore questionable	for	semiconductors excited by ultrafast laser pulses. 
			Comparison with carrier-phonon coupling	parameters extracted with the Boltzmann equation
			nevertheless shows, that the approximation is justified whenever the carrier system is
			not highly degenerate.\cite{Rethfeld2014}

			The lattice system is not directly heated up by the laser pulse but only indirectly via carrier-phonon coupling. Thus,
			the balance equation for lattice energy density reads
			\begin{align}
				\frac{\partial U_{ph}}{\partial t} &= \vec{\nabla}\cdot\left(\kappa_{ph}\,\vec{\nabla}T_{ph}\right) + g\,(T_e-T_{ph})\,,
				\label{eq:lattice_energy_balance}
			\end{align}
			where the first summand describes energy transport in the lattice system via Fourier's law and $\kappa_{ph}$ is the
			lattice thermal conductivity. The second term describes the energy exchange with the carrier system.
			This equation is basically identical to the one in the traditional TTM used for metals.\cite{Anisimov74}

			To calculate carrier and lattice temperature, we have to look at the temporal derivatives of carrier and lattice energy 
			densities in more detail. Lattice energy density only depends on lattice temperature.
			Thus, the temporal derivative of the lattice energy density can be expressed as
			\begin{align}
				\frac{\partial U_{ph}}{\partial t} = C_{ph}\,\frac{\partial T_{ph}}{\partial t}\,,
			\end{align}
			using the lattice heat capacity	$C_{ph}=\partial U_{ph}/\partial T_{ph}$.
			For the carrier system things are more complicated because the carrier energy density does not only depend on carrier
			temperature but also on carrier density and band gap energy:
			\begin{align}
				\frac{\partial U_{e-h}}{\partial t} = C_{e-h}\,\frac{\partial T_e}{\partial t} 
													+ \frac{\partial U_{e-h}}{\partial n_e}\,\frac{\partial n_e}{\partial t}
													+ \frac{\partial U_{e-h}}{\partial \varepsilon_g}\,\frac{\partial
													\varepsilon_g}{\partial t}\,.
				\label{eq:temporal_derivative_Ue}
			\end{align}
			Note that the temporal derivative of band gap energy is non-zero, because the band gap depends on both carrier
			density and lattice temperature.\cite{S-T00, Varshni67, vanDriel87}
			Furthermore, to calculate the derivatives of carrier energy	density, it is necessary to calculate the derivative of the reduced
			Fermi levels, as can be seen in Eq.~\eqref{eq:carrier_energy}. This is possible by taking the derivative with respect to
			carrier temperature, lattice temperature and carrier density, respectively, on both sides of
			Eq.~\eqref{eq:local_density}, utilizing that all three are independent properties and solving for the derivative 
			of the reduced Fermi level. 
			The reduced Fermi level itself can be calculated by solving Eq.~\eqref{eq:local_density} for the Fermi integral
			and comparing with tabulated data that can, for instance, be generated using the GNU Scientific Library.\cite{gsl}
		
		\subsection{Non-Degenerate Carrier System \label{sec:non-degenerate}}
			If the chemical potentials of electrons and holes are located far outside the conduction and the valence band,
			respectively, only the Boltzmann tail of the Fermi distribution reaches inside the band. The carrier system 
			is therefore non-degenerate	and follows a Maxwell-Boltzmann distribution. Consequently, the reduced Fermi levels 
			are large and negative and Fermi integrals of any order can be reduced to $\exp(\eta_e)$ and $\exp(\eta_h)$, 
			respectively. The quotients of Fermi integrals $\fermiquot{\xi}{\zeta}{\eta_c}$ tend towards one and the carrier 
			energy and pair current densities simplify from 
			Eqs.~\eqref{eq:carrier_pair_current_final} and \eqref{eq:ambipolar_energy_current} to
			\begin{align}
				\vec{J} = -{D}\,\left(\vec{\nabla}n_e + \frac{n_e}{2k_BT_e}\,\vec{\nabla}\varepsilon_g
				+\frac{n_e}{2T_e}\,\vec{\nabla}T_e\right)
				\label{eq:carrier_pair_current_non-deg}
			\end{align}
			with the ambipolar diffusivity ${D}=\frac{2k_BT_e}{e}\,\frac{\mu_e^0\mu_h^0}{\mu_e^0+\mu_h^0}$ and
			\begin{align}
				\vec{W} = \left(\varepsilon_g+4k_BT_e\right)\,\vec{J} - \left(\kappa_e+\kappa_h\right)\,\vec{\nabla}T_e\,.
				\label{eq:ambipolar_energy_current_non-deg}
			\end{align}

			The carrier energy density in a non-degenerate system is given by
			\begin{align}
				U_{e-h} = n_e\,\varepsilon_g + 3\,n_e\,k_B\,T_e\,,
				\label{eq:carrier_energy_non-deg}
			\end{align}
			which immensely simplifies the calculation of the carrier heat capacity and other derivatives in
			Eq.~\eqref{eq:temporal_derivative_Ue}. 
			
			Aside from the reduced Fermi levels being negative, another	criterion for whether the carrier system is degenerate 
			lies within the	comparison of the carrier temperature $T_e$	with the Fermi temperature
			\begin{align}
				T_F = \frac{\hbar^2}{2\,m^*_{c,\mathrm{DOS}}\,k_B}\,\left(3\,\pi^2\,n_e\right)^{2/3}\,,
				\label{eq:Fermi_temperature}
			\end{align}
			which is different for electrons and holes because of the different effective masses. 
			Whenever the Fermi temperature is comparable to or higher than the carrier temperature, the respective carrier 
			system is degenerate.
		
		\subsection{Laser-Excitation}
			To obtain the intensity of the laser pulse within the material, 
			the attenuation in the direction of propagation is calculated with the
			one-dimensional ordinary differential equation (ODE)
			\begin{align}
				\frac{\mathrm{d}I}{\mathrm{d}z}= -(\alpha_\mathrm{SPA}+\alpha_\mathrm{FCA})\,I - \beta_\mathrm{TPA} \,I^2\,,
				\label{eq:attenuation}
			\end{align}
			that includes
			linear interband or single photon absorption (SPA), linear intraband or free carrier absorption
			(FCA) and two photon interband absorption (TPA).
			Because of momentum conservation and the fact, that silicon is an indirect semiconductor, single photon
			processes can only happen under assistance of a phonon. Thus, the linear absorption coefficients
			$\alpha_\mathrm{SPA}$ and $\alpha_\mathrm{FCA}$ depend on lattice temperature. In addition, the 
			FCA coefficient depends on carrier density because 
			the more free carriers, the stronger the absorption.
			Because of the possibly steep carrier density and lattice temperature profiles within the material during
			irradiation, we have to account for	spatial non-constant absorption coefficients when solving the
			ODE~\eqref{eq:attenuation}.	Consequently, there is no closed-form analytical solution and the
			ODE~\eqref{eq:attenuation} has to be solved numerically along with the balance equations~\eqref{eq:particle_balance}, 
			\eqref{eq:carrier_energy_balance} and \eqref{eq:lattice_energy_balance}. 
			The intensity transmitted through the material surface is
			\begin{align}
				I_0 =\sqrt{\frac{4\,\ln(2)}{\pi}}\,\frac{(1-R)\,\Phi}{\tau_p}
						\,\exp\left\{-4\,\ln(2)\left[\frac{(t-t_0)}{\tau_p}\right]^2\right\}\,,
				\label{eq:surface_intensity}
			\end{align}
			where $R$ denotes the reflectivity, $\Phi$ the fluence and $\tau_p$ the duration of the laser pulse that
			is centered around $t_0=3\,\tau_p$.
				
			Because of the usually large spot size of the laser pulse, the radial intensity distribution can be neglected and it is
			sufficient to describe absorption and transport in the direction of propagation at the center of the focus. 

			As already indicated above, the huge changes in free carrier density during the irradiation lead to significant
			changes in the optical parameters, namely the reflectivity and the FCA coefficient.
			Nevertheless, when modeling semiconductors, the reflectivity is often assumed to depend solely on lattice
			temperature.\cite{vanDriel87,Chen05,Korfiatis07,Gan12} 
			In these cases the FCA coefficient is written as a solely lattice temperature dependent cross-section
			multiplied by carrier density. In the following, we will denote these expressions for the reflectivity and
			the FCA coefficient as $T$-expression.

			To account for the influence of the transient carrier density on reflectivity and FCA coefficient, we apply 
			a Drude model.
			Considering the separate contributions of both, electrons and holes, the complex dielectric function in the
			framework of this model is given by\cite{Gallant82, Callan00}
			\begin{align}
				\epsilon(\omega_L) 	&= \epsilon_r(\omega_L) - \frac{n_e\,e^2}{\epsilon_0\,\omega_L^2}\,
										\nonumber\\
									&\times
											\Bigg[\frac{1}{m_{e,\mathrm{cond.}}^*\,\left(1+i\,\frac{\nu_e}{\omega_L}\right)}
											+\frac{1}{m_{h,\mathrm{cond.}}^*\,\left(1+i\,\frac{\nu_h}{\omega_L}\right)}
											\Bigg]\,.
				\label{eq:dielectric_function}
			\end{align}
			Here, $\epsilon_r$ denotes the intrinsic dielectric constant, $m_{c,\mathrm{cond.}}^*$ is the conductivity
			effective mass defined to reproduce electrical conductivity and susceptibility.\cite{Spitzer57} 
			The Drude collision frequency of electrons and holes, respectively, is denoted $\nu_c$. 
			When assuming the collision frequencies of electrons and holes to be equal, $\nu_\mathrm{Drude}=\nu_e=\nu_h$, and 
			introducing a joint	optical effective mass, $1/m_\mathrm{opt}^*=1/m_{e,\mathrm{cond.}}^*+1/m_{h,\mathrm{cond.}}^*$, 
			Eq.~\eqref{eq:dielectric_function} reduces to the usual Drude expression for the dielectric function.\cite{Callan00,S-T00}

			The complex refractive index $\tilde{n}=\sqrt{\epsilon}$, the reflectivity under normal incidence and the FCA coefficient
			\begin{align}
				R=\frac{\left|\tilde{n}-1\right|^2}{\left|\tilde{n}+1\right|^2} 
				\qquad\text{and}\qquad
				\alpha_\mathrm{FCA} = \frac{2\,\mathfrak{Im}\left(\tilde{n}\right)\,\omega_L}{c_0}
			\label{eq:reflectivity_and_FCA_drude}
			\end{align}
			can subsequently be calculated using Eq.~\eqref{eq:dielectric_function}.\cite{fox} 
			Here, $c_0$ denotes the speed of light in vacuum.
			While the steep carrier density profile within the medium during excitation may cause feedback effects and thus
			influence the reflectivity of the silicon sample\cite{S-T00}, these effects are estimated as negligible for all
			situations modeled here.

			Concerning the collision frequency $\nu_c$, there is disagreement in literature on 
			the question which kinds of	collisions contribute (for an overview see review by Balling and Schou~\cite{Balling13}
			and references therein). Many authors assume a constant frequency 
			$\nu_\mathrm{Drude}$,\cite{Quere01,Rethfeld10CPP,S-T00,Silaeva13,Choi02,Zhang11,Gan11}
			while others consider non-constant electron-electron and electron-phonon 
			collision frequencies.\cite{Christensen09,Rethfeld10APA,Medvedev13CPP,Waedegaard13}

			Electron-electron as well as hole-hole collisions, however, cannot contribute to the collision frequency entering
			the Drude model as both particles have the same effective mass and the collision does conserve the total carrier 
			momentum.\cite{Preston84, Hulin84, Combescot87} Electron-hole collisions, on the other hand, can contribute to 
			the Drude collision frequency because of the different effective mass of both collision partners.
	
			Sernelius\cite{Sernelius89} as well as Hullin~\emph{et al.}\cite{Hulin84} investigated the importance of
			electron-hole collisions in comparison to carrier-phonon collisions and found that they can dominate under certain
			conditions. In the framework of the nTTM presented here, we thus consider both, electron-hole and carrier-phonon
			collisions.
			We assume the electron-phonon and the hole-phonon collision frequency to be identical and
			proportional to lattice temperature:\cite{Wang94}
			\begin{align}
				\nu_\mathrm{c-ph}=A\,T_{ph}\,.
				\label{eq:carrier-phonon_collision_frequency}
			\end{align}
			To our knowledge, there is no data for silicon or comparable semiconductors. For metals like gold, silver,
			copper or aluminum, Christensen \emph{et al.}\cite{Christensen07} state that the proportionality constant $A$
			is on the order of \unit{1\times10^{11}}{\reciprocal\second\reciprocal\kelvin} to 
			\unit{4\times10^{11}}{\reciprocal\second\reciprocal\kelvin}. Here, we assume that
			the proportionality constant for silicon is comparable to those for these metals and take
			$A=\unit{3.9\times10^{11}}{\reciprocal\second\reciprocal\kelvin}$ given for aluminum in
			Ref.~\onlinecite{Christensen07}.
			
			The electron-hole collision frequency is calculated using
			\begin{align}
				\nu_\mathrm{e-h} = \frac{\sqrt{\langle (\Delta v)^2\rangle}}{\langle\ell\rangle}\,,
				\label{eq:electron-hole_collision_frequency_general}
			\end{align}
			where $\langle\ell\rangle$ denotes the mean free path of electrons and holes for screened
			Coulomb collisions and 
			$\langle (\Delta v)^2\rangle=2\langle U_{e}^\mathrm{kin}\rangle/m_{e}^*+2\langle U_{h}^\mathrm{kin}\rangle/m_{h}^*$ 
			is the mean relative velocity squared of electrons and holes. 
			The mean electron and hole kinetic energies $\langle U_{c}^\mathrm{kin}\rangle$ can be calculated 
			by splitting the second summand in Eq.~\eqref{eq:carrier_energy}, that represents the carrier kinetic energy, into the 
			different parts for electrons and holes and dividing by carrier density. The mean free path 
			of electrons and holes in a screened Coulomb potential is given by\cite{Medvedev10}
			$\langle\ell\rangle=\varkappa^2/n_e\pi$, where the inverse screening length $\varkappa$ 
			for a free carrier gas can be calculated following Refs.~\onlinecite{DelFatti00,Binder97,Mueller13}:
			\begin{align}
				\varkappa^2 &= \frac{e^2}{\epsilon_0}\,\left[
									\int_{\varepsilon_C}^\infty\,\frac{\partial D_{e}(\varepsilon)}{\partial\varepsilon}\,f_{e}(\varepsilon)\,\mathrm{d}\varepsilon
									+ \int_{-\infty}^{\varepsilon_V}\,\frac{\partial D_{h}(\varepsilon)}{\partial\varepsilon}\,f_{h}(\varepsilon)\,\mathrm{d}\varepsilon
									\right]\nonumber\\
								&= \frac{e^2}{\epsilon_0}\,\frac{1}{\pi^{3/2}\hbar^3}\,\sqrt{\frac{k_BT_e}{2}}\nonumber\\
								&\quad \times	\left[\fermiint{-1/2}{\eta_e}\,m_{e,\mathrm{DOS}}^{*3/2} 
										+ \fermiint{-1/2}{\eta_h}\,m_{h,\mathrm{DOS}}^{*3/2}\right]\,.
								\label{eq:Coulomb_screening}
			\end{align}
			Note that this inverse screening length includes screening by both types of carriers, electrons and holes.
			Together with Eq.~\eqref{eq:electron-hole_collision_frequency_general} this leads to an electron-hole collision
			frequency of
			\begin{align}
				\nu_\mathrm{e-h}	&= \frac{\sqrt{3}\epsilon_0\pi\,(k_BT_e)^{3/2}}{e^2}\,
										\left[\frac{\fermiquot{3/2}{1/2}{\eta_e}}{m_{e,\text{DOS}}^*} 
											+ \frac{\fermiquot{3/2}{1/2}{\eta_h}}{m_{h,\text{DOS}}^*}
										\right]^{1/2}
										\nonumber\\
									&\quad
									\times\left[\fermiquot{-1/2}{1/2}{\eta_e}+\fermiquot{-1/2}{1/2}{\eta_h}\right]^{-1}\,.
				\label{eq:electron-hole_collision_frequency}
			\end{align}
			For a non-degenerate carrier system, this collision frequency reduces to
			\begin{align}
				\nu_\mathrm{e-h}    &= \frac{\sqrt{3}\epsilon_0\pi\,(k_BT_e)^{3/2}}{2e^2}
										\left[\frac{1}{m_{e,\text{DOS}}^*}+\frac{1}{m_{h,\text{DOS}}^*}\right]^{1/2}\,.
			\end{align}

			The total electron and hole collision frequencies can now be calculated from the respective single
			contributions \eqref{eq:carrier-phonon_collision_frequency} and \eqref{eq:electron-hole_collision_frequency} using
			Matthiessen's rule.
			Note that, following \eqref{eq:carrier-phonon_collision_frequency} and \eqref{eq:electron-hole_collision_frequency},
			the total electron and the total hole collision frequency are identical.

	\section{Results}
		The balance equations~\eqref{eq:particle_balance}, \eqref{eq:carrier_energy_balance} and
		\eqref{eq:lattice_energy_balance} are solved in one dimension using a finite differences scheme.
		While an explicit forward time centered space (FTCS) scheme is used to obtain lattice temperature and carrier density, 
		the	carrier temperature	is calculated using an iterative Crank Nicolson scheme.\cite{numerical_recipes} 
		The equations are solved on a staggered grid, where the grid points for carrier and energy currents 
		are positioned in between the grid points for temperatures and density. As initial
		conditions, we choose an equilibrium at $T_e(z,0)=T_{ph}(z,0)=\unit{300}{\kelvin}$
		and, consequently, $n_e(z,0)=\unit{10^{12}}{\centi\meter\rpcubed}$. 
		As energy-conserving boundary conditions, we use $J_z(z,t)=0$, $W_z(z,t)=0$ and 
		$\kappa_{ph}\,\partial T_{ph}/\partial z = 0$ 
		at the surface ($z=0$) and at $z=\unit{10}{\micro\meter}$ assumed as the thickness of the material. 
		The model parameters used for the calculations presented below are listed in Table~\ref{tab:model_parameters}. 
		It should be noted here that the nTTM like the commonly used TTM requires the use of various 
		material parameters as well as approximations. While some of these parameters are affected 
		with certain inaccuracies, numerous studies have proven the usefulness of the TTM in
		the past.\cite{Hohlfeld00, Bonn00, Vestentoft06, Christensen07, Byskov-Nielsen11}
		Obviously, it is possible to choose material parameters in such a way that the simulation 
		gives the best agreement with the experiment, e.g. by fitting certain material parameters.
		However, we would like to point out, that no fitting was performed in the present simulations. 
		The material parameters used here (cf. Tab.~\ref{tab:model_parameters}) are all commonly accepted 
		for the case of silicon. The sole exception is the carrier-phonon collision 
		frequency~\eqref{eq:carrier-phonon_collision_frequency}. 
		No value for the proportionality constant $A$ is know for silicon. The choice to use the value 
		for aluminum from Ref.~\onlinecite{Christensen07} is arbitrary and should not be understood as a fit.

		For all figures showing temporal evolutions, a \unit{100}{\femto\second}-laser pulse with a fluence of
		\unit{130}{\milli\joule\per\centi\meter\squared} at \unit{800}{\nano\meter} is applied.
		\begin{table*}
			\caption{\label{tab:model_parameters} Model parameters for silicon.}
			\begin{ruledtabular}
				\begin{tabular}{lclc}
					Quantity
						& Symbol
						& Value
						& Reference\\\hline
					\multicolumn{4}{c}{Band structure}\\
					Indirect band gap	
						& $\varepsilon_g$	
						& $\left(1.16-\frac{7.02\times10^{-4}\,\reciprocal\kelvin\,T_{ph}^2}{T_{ph}+1108\,\kelvin} - 1.5\times10^{-8}\,\centi\meter\,n_e^{1/3}\right)\,\electronvolt$
						& Refs.~\onlinecite{vanDriel87,Varshni67,S-T00}\\
					Electron DOS effective mass
						& $m^*_{e,\mathrm{DOS}}$
						& $0.33\,m_e$
						&Ref.~\onlinecite{Chen05}\\
					Hole DOS effective mass
						& $m^*_{h,\mathrm{DOS}}$
						& $0.81\,m_e$
						&Ref.~\onlinecite{Barber67}\\
					\multicolumn{4}{c}{}\\
					\multicolumn{4}{c}{Thermal and electrical properties}\\
					Electron conductivity effective mass
						& $m^*_{e,\mathrm{cond.}}$
						& $0.26\,m_e$
						& Ref.~\onlinecite{vanDriel87}\\
					Hole conductivity effective mass
						& $m^*_{h,\mathrm{cond.}}$
						& $0.37\,m_e$
						& Ref.~\onlinecite{vanDriel87}\\
					Lattice heat capacity
						& $C_{ph}$
						& $\left(1.978+3.54\times10^{-4}\,\reciprocal\kelvin\,T_{ph} - 3.68\,\kelvin\squared\,T_{ph}^{-2}\right)\,\joule\per(\centi\meter\cubed\,\kelvin)$
						& Refs.~\onlinecite{vanDriel87,Wood81}\\
					Lattice thermal conductivity
						& $\kappa_{ph}$
						& $1585\,\kelvin^{1.23}\,T_{ph}^{-1.23}\,\watt\per(\centi\meter\,\kelvin)$
						& Refs.~\onlinecite{vanDriel87,Wood81}\\
					Melting temperature
						& $T_{\rm melt}$
						& \unit{1685}{\kelvin}
						& Ref.~\onlinecite{S-T00}\\
					Carrier thermal conductivity
						& $\kappa_{\rm e-h}$
						& $\left(-3.47\times10^{16} + 4.45\times10^{14}\,\reciprocal\kelvin\,T_e\right)\,\electronvolt\per(\second\,\centi\meter\,\kelvin)$
						& Refs.~\onlinecite{Chen05,Agassi84}\\
					Carrier ambipolar diffusivity
						& $D$
						& $18\,(300\,\kelvin/T_{ph})\,\centi\meter\squared\per\second$
						& Ref.~\onlinecite{vanDriel87}\\
					Auger recombination coefficient
						& $\gamma$
						& $3.8\times10^{-31}\,\centi\meter^6\per\second$
						& Refs.~\onlinecite{vanDriel87, Dziewior77}\\
					Impact ionization coefficient
						& $\delta$
						& $3.6\times10^{10}\,\exp[-1.5\,\varepsilon_g/(k_BT_e)]\,\reciprocal\second$
						& Ref.~\onlinecite{vanDriel87}\\
					Carrier-phonon relaxation time
						& $\tau_{\rm c-ph}^{\rm relax}$
						& \unit{0.5}{\pico\second}
						& Ref.~\onlinecite{vanDriel87}\\
					\multicolumn{4}{c}{}\\
					\multicolumn{4}{c}{Optical properties at \unit{800}{\nano\meter}}\\
					Single photon absorption coefficient
						& $\alpha_{\rm SPA}$
						& $1.12\times10^{3}\,\exp(T_{ph}/430\,\kelvin)\,\centi\reciprocal\meter$
						& Ref.~\onlinecite{Korfiatis07}\\
					Two photon absorption coefficient
						& $\beta_\mathrm{TPA} $
						& \unit{9}{\centi\meter\per\giga\watt}
						& Ref.~\onlinecite{Korfiatis07}\\
					\multicolumn{4}{c}{}\\
					\multicolumn{4}{c}{$T$-expression}\\
					Free carrier absorption coefficient
						& $\alpha_{\rm FCA}$
						& $2.56\times10^{-18}\,\centi\meter\squared\,(T_{ph}/300\,\kelvin)\,n_e$
						& Ref.~\onlinecite{Korfiatis07}\\
					Reflectivity
						& $R$
						& $0.329 + 5\times10^{-5}\,\reciprocal\kelvin\,(T_{ph}-300\,\kelvin)$
						& Ref.~\onlinecite{Korfiatis07}\\
					\multicolumn{4}{c}{}\\
					\multicolumn{4}{c}{Drude model}\\
					Free carrier absorption coefficient
						& $\alpha_{\rm FCA}$
						& calculated using Eq.~\eqref{eq:reflectivity_and_FCA_drude}
						& Ref.~\onlinecite{fox}\\
					Reflectivity
						& $R$
						& calculated using Eq.~\eqref{eq:reflectivity_and_FCA_drude}
						& Ref.~\onlinecite{fox}\\
					Intrinsic dielectric constant
						& $\epsilon_r$
						& $13.634+0.048\,i$
						& Ref.~\onlinecite{palik}
				\end{tabular}
			\end{ruledtabular}
		\end{table*}

\begin{figure}
			\includegraphics{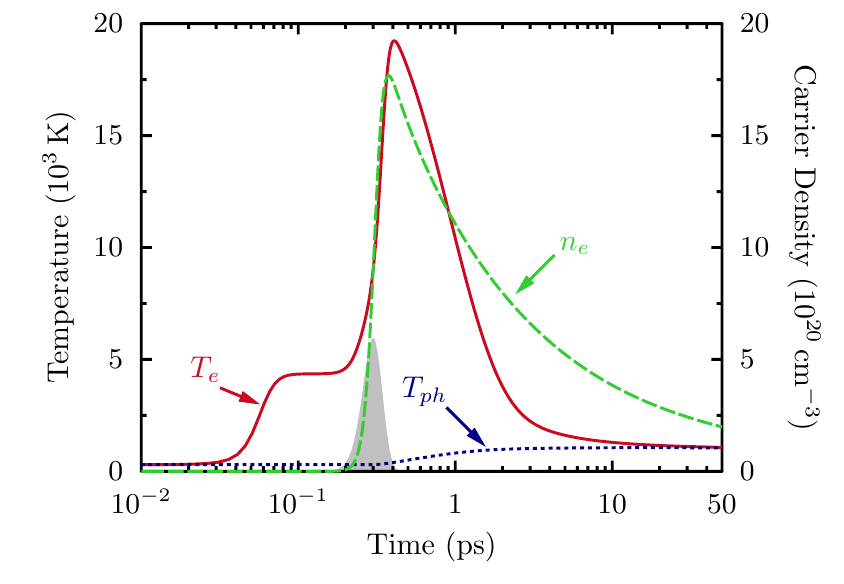}
			\caption{\label{fig:default} (Color online) Characteristic temperature and density evolution at
				the surface	of laser-irradiated silicon obtained with the nTTM.
				The laser intensity is sketched in arbitrary units as a gray area.}
		\end{figure}

		Figure~\ref{fig:default} shows the temporal evolution of carrier density (green dashed curve)
		and temperature (red solid curve) as well as lattice temperature (blue dotted curve) at the surface of laser-irradiated 
		silicon. 
		The data presented in the figure was calculated using the full model considering the transient optical parameters
		obtained with the Drude model, a Fermi distributed carrier system and transport.
		
		The carrier temperature starts to increase early at very low intensities until it reaches a	plateau of approximately 
		\unit{4360}{\kelvin} at about \unit{90}{\femto\second}. 
		The first increase is due to linear absorption processes, to be specific SPA and to some extend FCA. 
		The excess energy gained by these processes, namely all the energy that is not invested in overcoming the band
		gap, corresponds to a quasi-temperature of approximately \unit{4380}{\kelvin}. 
		Due to carrier-phonon coupling the actual carrier temperature is slightly lower.

		In the following \unit{100}{\femto\second},	carrier temperature only increases slightly before it ascends rapidly when 
		the laser pulse approaches its maximum. At this point, on the one hand, TPA becomes important for the high intensities 
		near the maximum of the pulse and, on the other hand, FCA becomes the dominant linear absorption process because of an 
		increasing free carrier density.

		The carrier temperature reaches its maximum slightly after the maximum of the laser pulse and then decreases mainly
		because of lattice heating.

		Due to carrier-phonon coupling, carrier and lattice temperatures tend to equilibrate. Consequently, the lattice
		heats up at later times than the carriers. Total equilibration is not achieved for several tens of picoseconds because,
		even on this timescale,	the carrier system is still heated up slightly as a result of Auger recombination.

		The carrier density reaches its maximum shortly after the laser pulse and slightly earlier than carrier temperature. The
		maximum carrier density of 
		\unit{1.77\times10^{21}}{\centi\meter\rpcubed} 
		is nine orders of magnitude larger than the initial
		carrier density. The increase is mainly due to photon absorption, while impact ionization appears 
		to be insignificant. After reaching its maximum, the free carrier density starts to decrease due to Auger recombination.

		In the following subsections, we will investigate the influence of several processes and properties on the outcome of
		our simulations. To further check the importance of the process or property under investigation and at the same time
		validate our model against the experiment, we present and discuss calculated damage thresholds at the end of each subsection. 
		As a criterion for damage, we choose that the maximum lattice temperature reaches melting temperature
		$T_\mathrm{melt}=\unit{1685}{\kelvin}$ as this represents the lowest fluence, at which damage can possibly occur. 
		The damage threshold curve for the full model as
		well as experimental data published by Allenspacher~\emph{et al.}\cite{Allenspacher03} and Pronko~\emph{et
		al.}\cite{Pronko98} is always shown as a reference.
		
		\subsection{Influence of Density-Dependent Optical Parameters\label{sec:implementing_drude}}
			In the last section, we demonstrated that the free carrier density in silicon may increase by
			as much as nine orders of magnitude
			during the irradiation with an ultrashort laser pulse. In the following, we will investigate the importance of
			the transient, density-dependent optical parameters. 
			For simplicity, we assume here that the carrier system is non-degenerate and neglect carrier and energy transport.
			
			\begin{figure}
				\subfloat{\label{fig:drude_vs_T-expression_t-R}}
				\subfloat{\label{fig:drude_vs_T-expression_t-n-Te-Tp}}
				\includegraphics{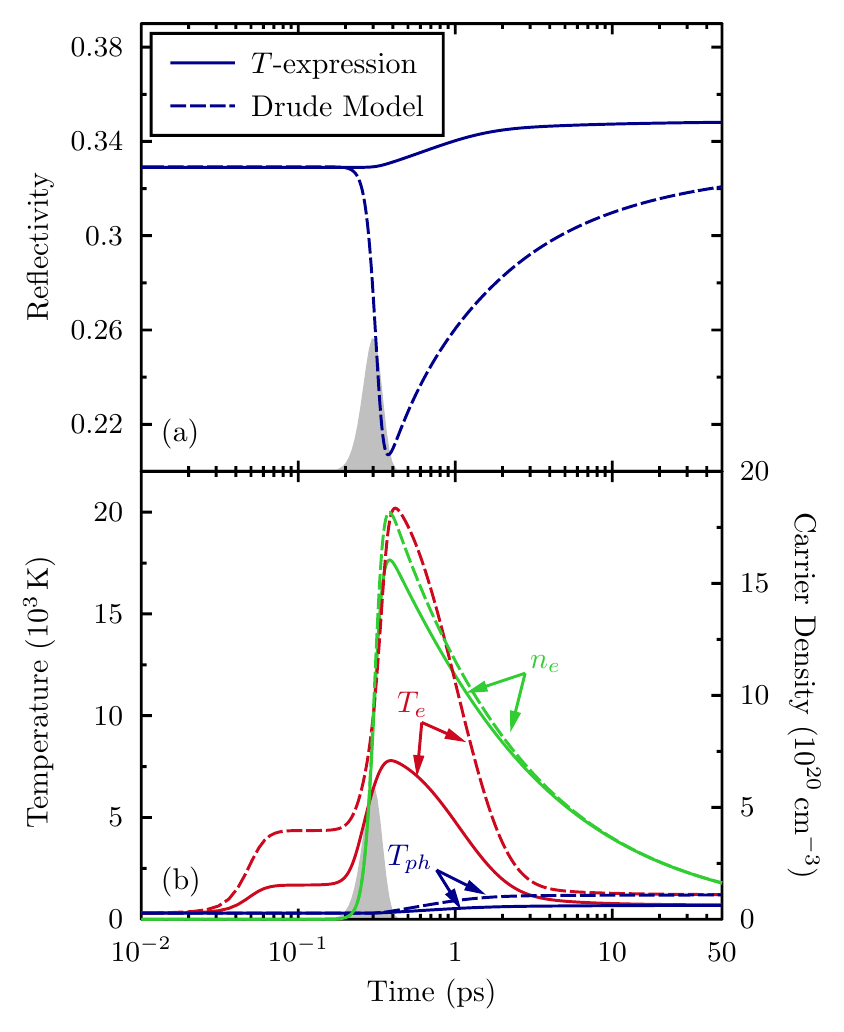}
				\caption{\label{fig:drude_vs_T-expression} (Color online) Reflectivity (a), carrier density as well as carrier and
				lattice temperature (b) calculated using either the $T$-expression or the Drude model.}
			\end{figure}
			Figure~\ref{fig:drude_vs_T-expression}\subref{fig:drude_vs_T-expression_t-R} depicts the reflectivity during the irradiation 
			of silicon calculated using	the $T$-expression (solid curve) and the Drude model (dashed curve), respectively, to determine
			both the reflectivity and the FCA coefficient as denoted in Tab.~\ref{tab:model_parameters}. We immediately notice major 
			differences: While the reflectivity
			calculated with the $T$-expression increases mirroring the behavior of lattice temperature, the reflectivity calculated
			with the Drude model decreases mirroring the inverse behavior of carrier density. The reflectivity calculated with the 
			Drude model already shows significant changes during irradiation, while the increase calculated with the $T$-expression 
			is delayed and less pronounced. 

			Figure~\ref{fig:drude_vs_T-expression}\subref{fig:drude_vs_T-expression_t-n-Te-Tp} clearly shows that the
			drastically different behavior of reflectivity has an impact on carrier and lattice temperatures as well as carrier density. 
			The maxima of carrier density as well as carrier and lattice temperatures are all higher when using the Drude model than 
			when using the $T$-expression. This is most prominent for carrier temperature but also noticeable for the final lattice 
			temperature. The densities calculated using the Drude model and the $T$-expression, respectively, deviate at their maximum 
			but tend towards similar values for later times. The reason is that Auger recombination, as a three particle process, is 
			much stronger for higher densities.
			
			The significant differences, especially in temperature, shown in
			Fig.~\ref{fig:drude_vs_T-expression}\subref{fig:drude_vs_T-expression_t-n-Te-Tp} indicate that the choice of using
			either the $T$-expression or the Drude model to calculate reflectivity and FCA coefficient might strongly
			influence the damage thresholds estimated applying the nTTM. 

			\begin{figure}
				\includegraphics{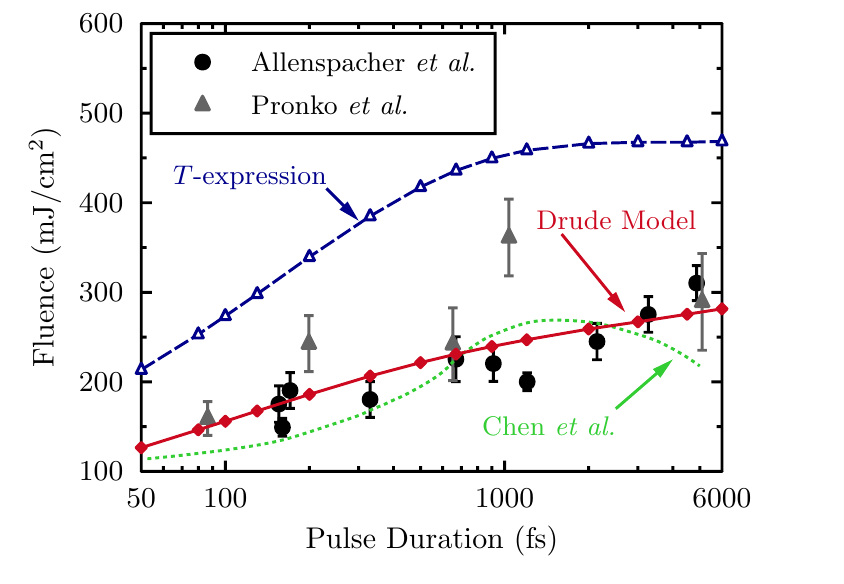}
				\caption{\label{fig:threshols_drude_vs_T} (Color online) Damage thresholds calculated using either the
				T-expression or the full model and thereby the Drude model in comparison to experimental 
				data obtained by Allenspacher~\emph{et al.}\cite{Allenspacher03} and Pronko~\emph{et al.}\cite{Pronko98} as well 
				as thresholds simulated by Chen~\emph{et al.}\cite{Chen05}}
			\end{figure}

			In Fig.~\ref{fig:threshols_drude_vs_T} the damage thresholds calculated using the $T$-expression (blue triangles)
			and the	full model including density-dependent optical parameters (red diamonds) are shown in comparison to 
			experimental data published by Allenspacher~\emph{et al.}\cite{Allenspacher03} and Pronko~\emph{et al.}\cite{Pronko98} 
			as well as theoretical predictions presented by Chen~\emph{et al.}\cite{Chen05}. 
			Both our calculations consider transport and a Fermi distributed carrier gas. For their	calculations, Chen~\emph{et
			al.}\cite{Chen05} solved an nTTM in combination with the $T$-expression. As damage criterion, they used a critical 
			density fitted to match the experimental determined threshold for a pulse duration of \unit{500}{\femto\second}.

			Figure~\ref{fig:threshols_drude_vs_T} clearly shows, that our calculation using the $T$-expression	significantly 
			overestimates the damage thresholds while the damage thresholds calculated using the Drude model are in very good 
			agreement with the experimental data even over a wide range of pulse durations from \unit{80}{\femto\second} 
			up to \unit{6}{\pico\second}. 
			The thresholds calculated by Chen~\emph{et al.}\cite{Chen05} agree reasonably well with the experiment. However, 
			the threshold decreases for long pulses, whereas the experimental data does not show any decrease. For these pulse 
			durations in the picosecond range, the thresholds calculated with our full model considering density-dependent optical
			parameters much	better resemble the behavior of the experimental data. 

			We therefore conclude, that it is highly important to account for the transient optical parameters changing due to the
			strongly non-constant free carrier density in laser-excited silicon.
			
		\subsection{Influence of Collision Frequency\label{sec:influence_collfreq}}
			Following this conclusion, 
			we will now look at the reflectivity obtained with the Drude model, especially the importance 
			of the choice of carrier collision frequency, in more detail.
			We will again neglect transport and assume the carrier system to be non-degenerate at all times.
		
			\begin{figure}[t]
				\subfloat{\label{fig:drude_influence_of_frequency_t-R}}
				\subfloat{\label{fig:drude_influence_of_frequency_t-FCA}}
				\includegraphics{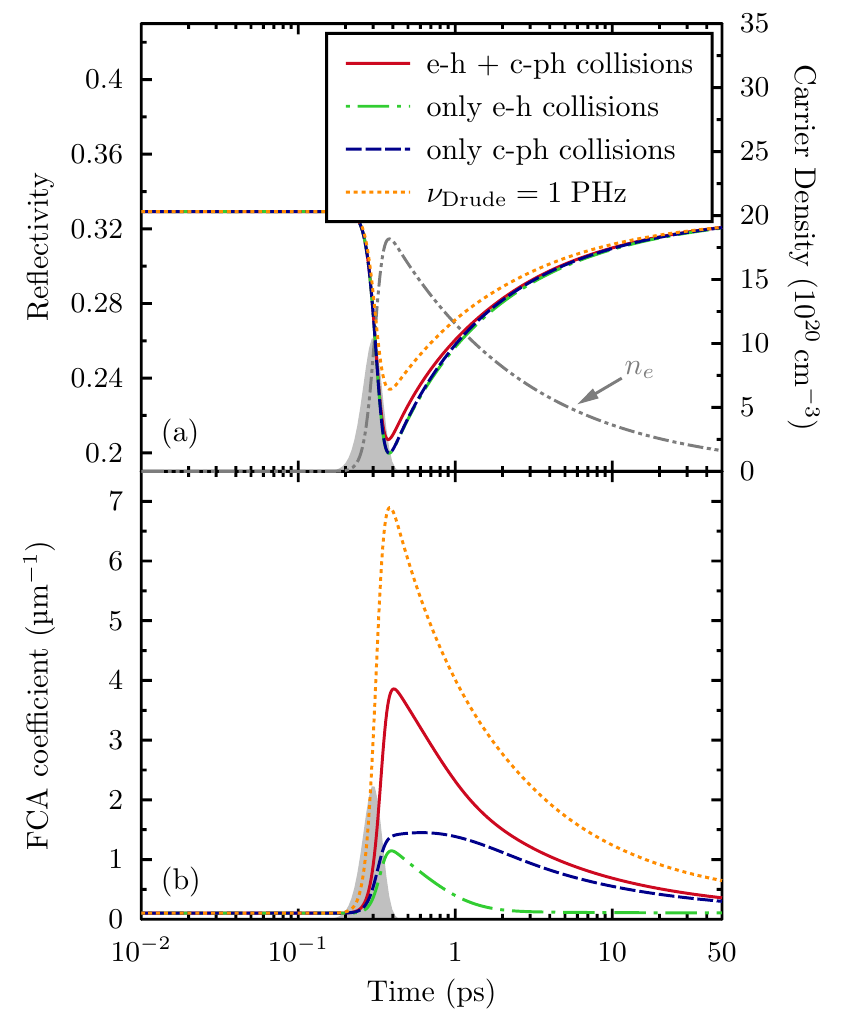}
				\caption{\label{fig:drude_influence_of_frequency_t-R-FCA} (Color online) Reflectivity (a) and FCA coefficient (b) 
				assuming a constant frequency of \unit{1}{\peta\hertz}, considering either electron-hole collisions or 
				carrier-phonon collisions or considering both, electron-hole and carrier-phonon collisions. As an orientation, 
				the carrier density considering both types of collisions is shown as a gray dash-dot-dotted 
				line. Note that the carrier density is nearly identical for all carrier collision frequencies shown here.}
			\end{figure}
			Figure~\ref{fig:drude_influence_of_frequency_t-R-FCA}\subref{fig:drude_influence_of_frequency_t-R} depicts the
			reflectivity calculated with the Drude model for different assumptions on the collision frequency. 
			The reflectivity considering 
			electron-hole 
			and carrier-phonon collisions (red solid curve) is the same as shown in
			Fig.~\ref{fig:drude_vs_T-expression}\subref{fig:drude_vs_T-expression_t-R}. 
			In this case, the total electron and hole collision frequency varies from \unit{118}{\tera\hertz}
			to maximum \unit{500}{\tera\hertz}. 
			While the carrier-phonon collisions dominate the total collision frequency in the beginning for low intensities 
			and later after the pulse, the 
			electron-hole collisions 
			dominate near the maximum of the laser pulse.

			To investigate, whether the contribution of any type of collision to the Drude collision frequency can be
			neglected or it is even sufficient to assume a constant collision frequency,
			Fig.~\ref{fig:drude_influence_of_frequency_t-R-FCA}\subref{fig:drude_influence_of_frequency_t-R} additionally
			shows the reflectivity considering only	electron-hole 
			(green dash-dotted curve) or carrier-phonon collisions (blue
			dashed curve) as well as the reflectivity for a constant collision frequency of\cite{S-T00} \unit{1}{\peta\hertz}
			(orange dotted curve).

			In all cases, the reflectivity follows the inverse behavior of the carrier density (gray dash-dot-dotted curve).
			The depths of the minima, however, are different. 
			The reflectivity curves considering only electron-hole 
			or carrier-phonon collisions are nearly identical and show the deepest
			minimum, while the reflectivity for a constant collision frequency shows a less pronounced minimum.

			The second optical property, that is directly influenced by the choice of the collision frequency, is the FCA
			coefficient. Figure~\ref{fig:drude_influence_of_frequency_t-R-FCA}\subref{fig:drude_influence_of_frequency_t-FCA}
			shows that all FCA coefficients roughly	follow the behavior of carrier density. It is, however, obvious that
			the strength of FCA differs depending on the choice of the collision frequency. For the carrier density range considered
			here ($n_e\le\unit{7\times10^{21}}{\centi\meter\rpcubed}$), the FCA coefficient predicted by the Drude model is higher for 
			a higher collision frequency. This is directly reflected in the figure: the constant collision frequency of
			\unit{1}{\peta\hertz} is higher than the collision frequency considering both types of collisions at all times, as
			is the FCA coefficient. 
			When only electron-hole 
			or carrier-phonon collisions are considered, the collision frequency
			is always lower than when both are considered which is directly reflected in the corresponding FCA coefficients.

			Comparing the FCA coefficients considering only one type of collision, we see that while the one for electron-hole
			collisions decreases rapidly after the end of the laser pulse, the one for carrier-phonon collisions decreases much slower.
			This is because electron-hole collisions are mainly important at the maximum of the pulse,
			when the carrier temperature is highest, while carrier-phonon collisions stay important once the lattice
			temperature increases due to carrier-phonon coupling.

			\begin{figure}
				\subfloat{\label{fig:drude_influence_of_frequency_t-Te-Tph_a}}
				\subfloat{\label{fig:drude_influence_of_frequency_t-Te-Tph_b}}
				\includegraphics{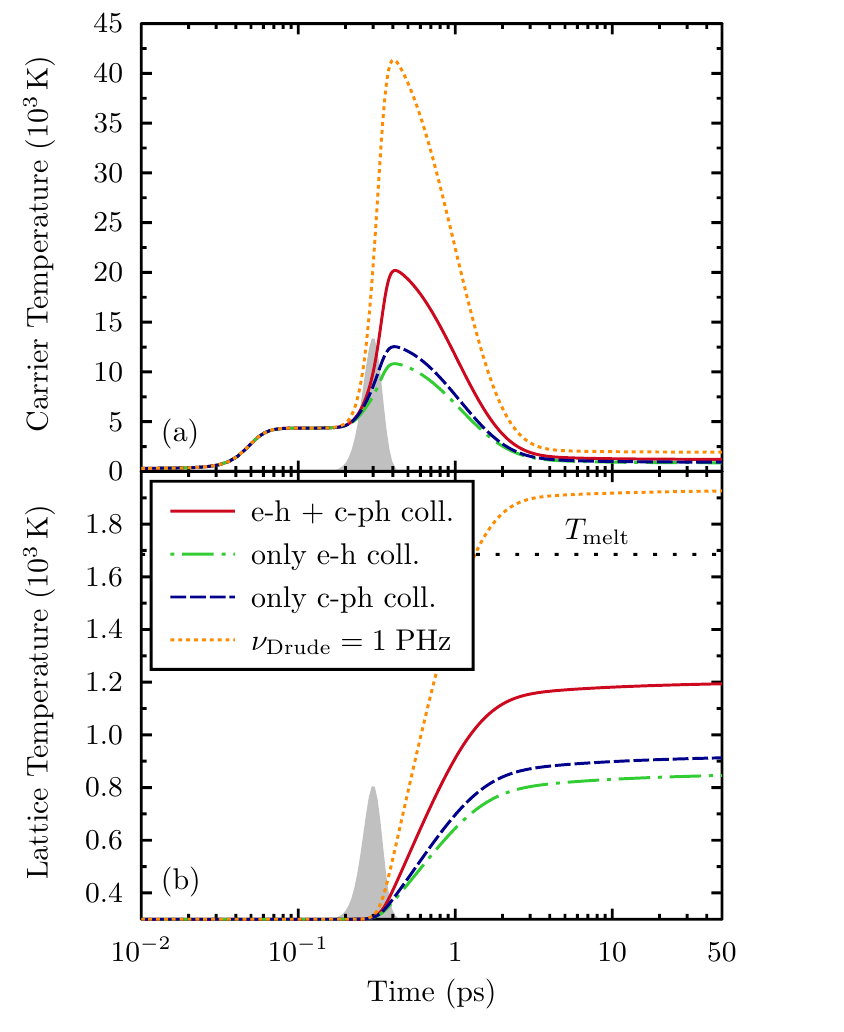}
				\caption{\label{fig:drude_influence_of_frequency_t-Te-Tph} (Color online) Influence of the carrier collision frequency used in
				the Drude model on carrier (a) and lattice temperature (b). Carrier density is not shown because there are only minor
				differences.}
			\end{figure}
			To further investigate the importance of the choice of the collision frequency,
			Fig.~\ref{fig:drude_influence_of_frequency_t-Te-Tph} shows carrier and phonon temperature for the different
			collision frequencies. Note that the carrier density is not shown here, because it is nearly identical in all
			cases. This can, however, not be said for the temperatures. The strongly different FCA coefficients shown in
			Fig.~\ref{fig:drude_influence_of_frequency_t-R-FCA}\subref{fig:drude_influence_of_frequency_t-FCA} lead to
			strongly different surface temperatures. 
			While the highest lattice temperature, obtained for the constant frequency,
			is already above melting temperature, the lowest final temperatures, when only one type of collision is considered 
			are	between 
			\unit{840}{\kelvin} and \unit{920}{\kelvin}.

			We conclude, that the calculated temperatures depend highly on the assumption on carrier collision frequency.
			Consequently, this assumption might significantly alter the prediction of damage thresholds. Here, we will
			compare the damage threshold when considering both types of collisions with thresholds obtained for constant collision 
			frequencies.

			\begin{figure}
				\includegraphics{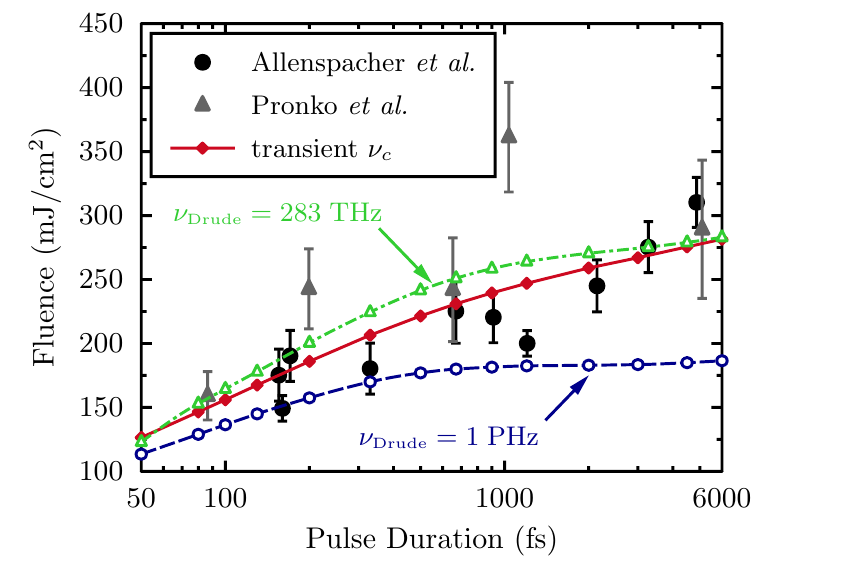}
				\caption{\label{fig:threshold_influence_frequency} (Color online) Damage thresholds calculated using 
				the full model with a transient carrier collision frequency as well as two different constant collision 
				frequencies in comparison to experimental data published by 
				Allenspacher~\emph{et al.}\cite{Allenspacher03} and Pronko~\emph{et al.}\cite{Pronko98}}
			\end{figure}

			Figure~\ref{fig:threshold_influence_frequency} depicts the damage thresholds calculated with the full 
			model and, hence, a	transient carrier collision frequency
			(solid red diamonds, same curve	appears in Fig.~\ref{fig:threshols_drude_vs_T})
			as well as with	two different constant collision frequencies. The collision frequency of \unit{1}{\peta\hertz} 
			(blue circles) can be found in literature\cite{S-T00}, while the frequency of \unit{283}{\tera\hertz} 
			(green triangles) was determined by fitting to an experimental data point by 
			Allenspacher~\emph{et al.}\cite{Allenspacher03} for a pulse duration in the picosecond range.

			For short pulses with durations up to about \unit{300}{\femto\second}, all calculated damage thresholds are 
			in good agreement with the experimental data. For longer pulses, however, the threshold curves for constant
			carrier collision frequencies flatten while the damage threshold for a transient frequency continues to increase.
			Because of the flattening in the threshold curve, the calculation using a constant collision frequency of
			\unit{1}{\peta\hertz} underestimates the damage threshold for picosecond pulses. 
			While calculations using the lower frequency of \unit{283}{\tera\hertz} predict fluences that compare very well
			with experimental results even for picosecond pulses, the further increase in the damage threshold for a 
			transient collision frequency best resembles the experimental data.
			This increase is due to the fact, that for longer pulses the maximum carrier temperature and thus, 
			according to Eq.~\eqref{eq:electron-hole_collision_frequency}, the transient 
			electron-hole 
			collision frequency decrease. Consequently, FCA becomes less effective for longer 
			pulse durations which causes the threshold for a transient frequency to increase further. 
			
			In summary, we conclude	that all thresholds calculated here agree reasonably well with the
			experimental results for short pulse durations. The threshold calculated using a transient collision 
			frequency is, however, the one that best reproduces the overall behavior indicated by the experimental data. 
			Moreover, no fitting is involved in generating this threshold curve, while it is necessary to fit the constant
			frequency to achieve an reasonable agreement with the experimental data over a range of pulse durations. 
			Consequently, a transient collision frequency should be applied if a wide range of pulse durations or 
			picosecond pulses are to be treated.

		\subsection{Influence of Degeneracy \label{sec:influence_of_degeneracy}}
			As has been discussed in Sec.~\ref{sec:non-degenerate}, the system of equations drastically simplifies when the carrier
			system is assumed to be non-degenerate (Maxwell-Boltzmann distributed). In the following, we will investigate, 
			whether this assumption is justified in laser-excited silicon. For that purpose, we use a transient carrier
			collision frequency and again neglect transport.

			The distribution functions of electrons and holes enter the equation for carrier energy density~\eqref{eq:carrier_energy}
			and consequently the calculation of carrier temperature. Thus, when abandoning
			the assumption of a classical non-degenerate carrier gas, the expression for carrier energy~\eqref{eq:carrier_energy_non-deg} 
			becomes	invalid and the full equation~\eqref{eq:carrier_energy} has to be used instead. Consequently, 
			the calculation of carrier temperature following Eq.~\eqref{eq:temporal_derivative_Ue} 
			also becomes more laborious.

\begin{figure}
				\subfloat{\label{fig:influence_of_degenerate_gas_on_n_T_a}}
				\subfloat{\label{fig:influence_of_degenerate_gas_on_n_T_b}}
				\includegraphics{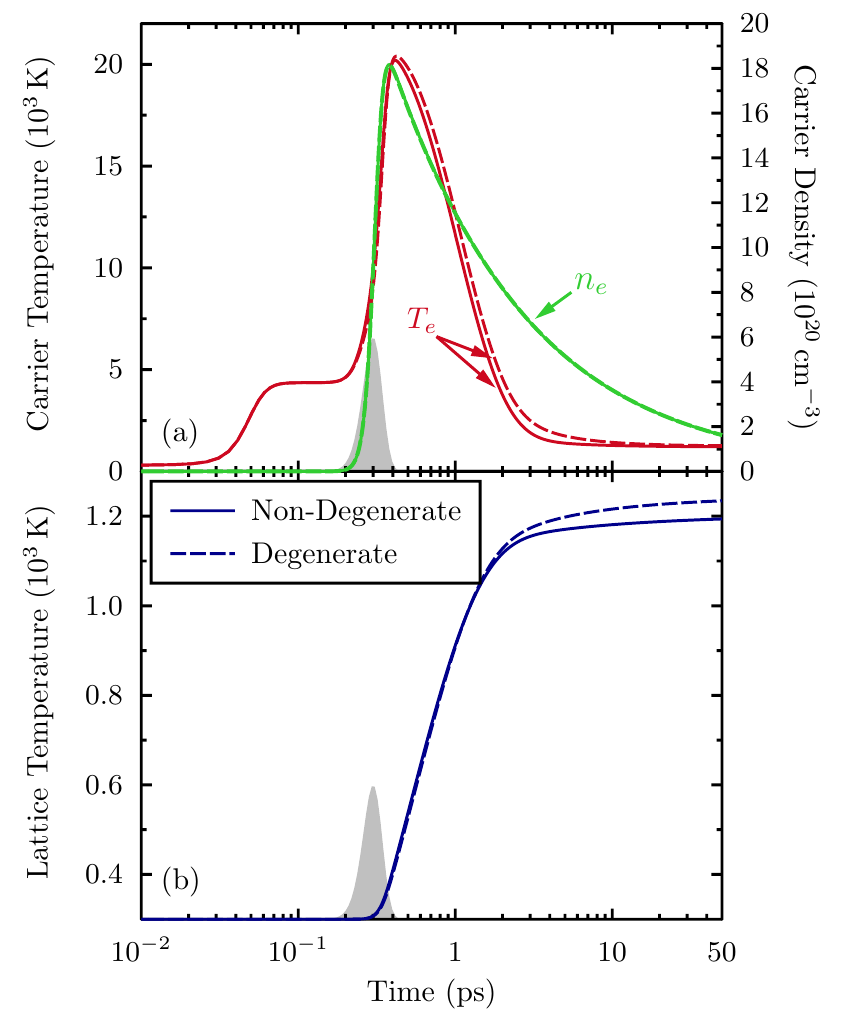}
				\caption{\label{fig:influence_of_degenerate_gas_on_n_T} (Color online) Carrier density and temperature (a) as well
				as lattice temperature (b) calculated assuming either a Maxwell-Boltzmann distributed or a Fermi distributed
				carrier system.}
			\end{figure}

			Figure~\ref{fig:influence_of_degenerate_gas_on_n_T} shows carrier densities and temperatures as well as	lattice
			temperatures calculated assuming a non-degenerate carrier gas and allowing the carrier gas to become degenerate,
			respectively.
			While the choice of carrier distribution function has almost no influence on carrier density, there are differences 
			in carrier temperature and, especially, lattice temperature
			(Fig.~\ref{fig:influence_of_degenerate_gas_on_n_T}\subref{fig:influence_of_degenerate_gas_on_n_T_b}).

			\begin{figure}
				\subfloat{\label{fig:comparison_Te_TF_a}}
				\subfloat{\label{fig:comparison_Te_TF_b}}
				\includegraphics{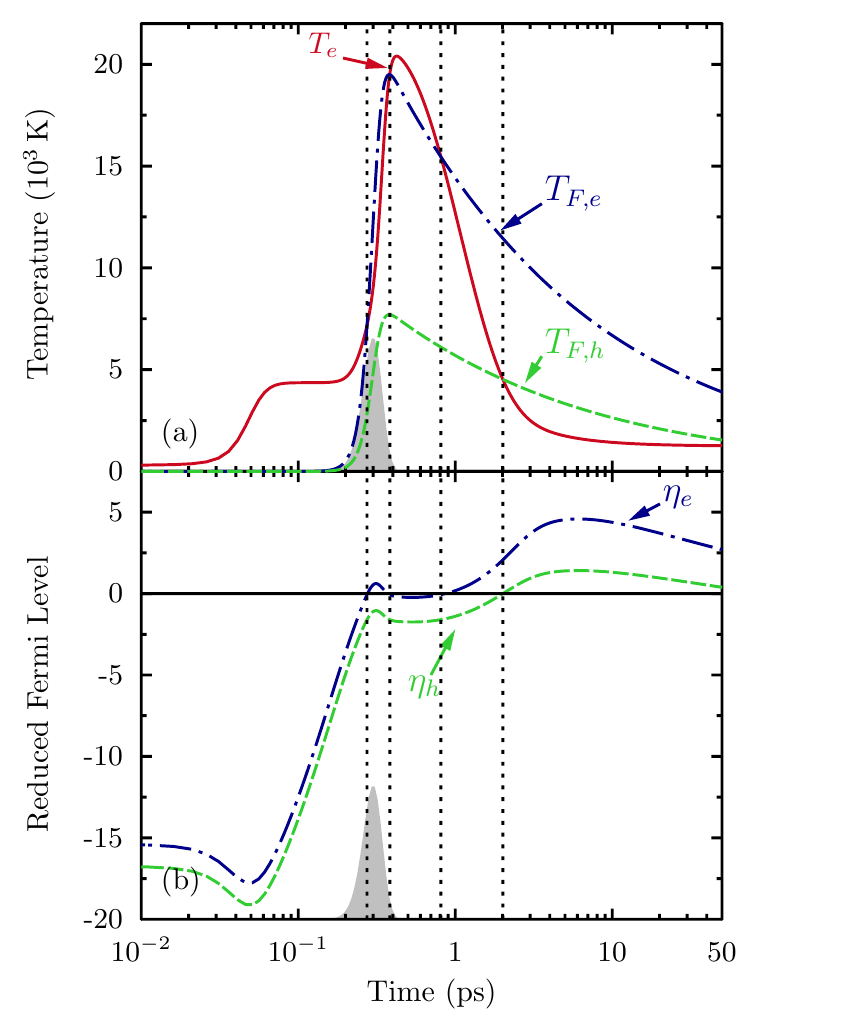}
				\caption{\label{fig:comparison_Te_TF} (Color online) Carrier temperature in comparison with Fermi
				temperature for electrons and holes, respectively, (a) and reduced Fermi levels of electrons and holes (b).}
			\end{figure}
			To check whether and when the carrier system becomes degenerate, carrier temperature and Fermi temperatures of
			electrons and holes \eqref{eq:Fermi_temperature} are depicted in
			Fig.~\ref{fig:comparison_Te_TF}\subref{fig:comparison_Te_TF_a}. If the Fermi temperature of electrons and holes is
			larger than carrier temperature, the respective carrier system is degenerate and should
			be described by a Fermi distribution. While the electron system is degenerate for a short time
			during the laser pulse and then again after the laser pulse, the hole system only becomes degenerate after the pulse
			at later times than the electrons. 

			Alternatively, we can check whether the chemical potential of the electrons (holes) is positioned inside the
			conduction (valence) band, which is another indication for degeneracy.
			It is equivalent to	the question whether the reduced Fermi levels of electrons and holes \eqref{eq:reduced_fermi_level} 
			are negative (chemical potential is positioned outside the band) or positive (chemical potential is
			positioned inside the band). Figure~\ref{fig:comparison_Te_TF}\subref{fig:comparison_Te_TF_b} shows the reduced Fermi
			levels of electrons and holes. It is obvious that this criterion indicates a degenerate electron and hole system in the
			same time intervals as the comparison with Fermi temperature. 
			In addition, we see that the degeneracy of electrons is more pronounced than the degeneracy of holes because the reduced
			Fermi level of the electrons is larger due to different effective masses. This means in turn that the chemical potential 
			of the electrons is	positioned deeper within the conduction band than the one of the holes is in the valence band.

			\begin{figure}
				\includegraphics{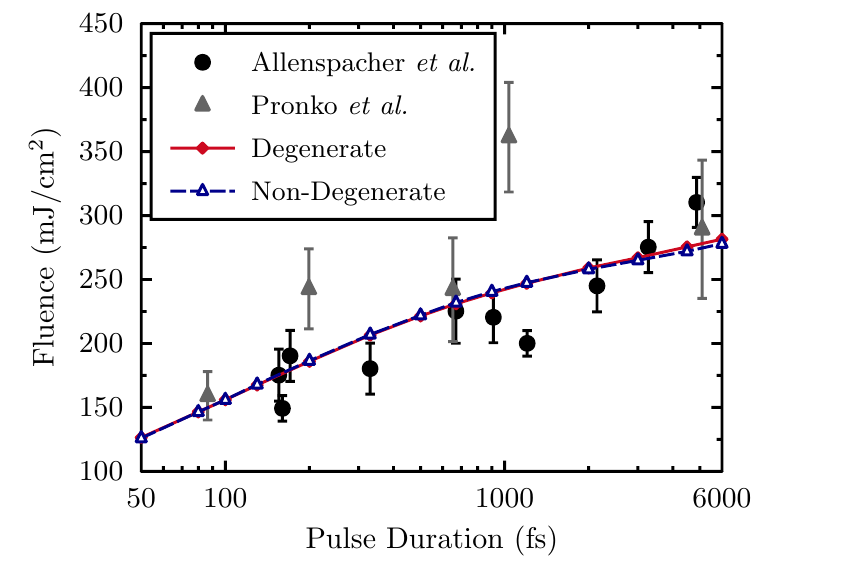}
				\caption{\label{fig:threshold_influence_degeneracy} (Color online) Damage threshold calculated
				assuming either a Maxwell-Boltzmann distribution or using the full model considering a
				Fermi distributed carrier system in comparison to experimental data
				published by Allenspacher~\emph{et al.}\cite{Allenspacher03} and Pronko~\emph{et al.}\cite{Pronko98}}
			\end{figure}
			
			The predicted damage thresholds, shown in Fig.~\ref{fig:threshold_influence_degeneracy}, 
			are nearly identical and are both in very good agreement with the experimental data.

			We conclude that though the electron as well as the hole system become degenerate during and after irradiation, it
			suffices to apply a Maxwell-Boltzmann distribution and, thus, to consider a non-degenerate carrier system, when 
			estimating damage thresholds. This immensely reduces computational effort.

		\subsection{Influence of Transport}
			In this section, we will investigate transport effects. 
			To that end, we consider both particle
			\eqref{eq:carrier_pair_current_final} and energy \eqref{eq:ambipolar_energy_current} transport.
			In all following calculations, the full system of equations is solved, thus allowing for degeneracy of the electron
			and hole system, respectively.
			\begin{figure}
				\subfloat{\label{fig:transport_surface_t-n-Te-Tp_a}}
				\subfloat{\label{fig:transport_surface_t-n-Te-Tp_b}}
				\includegraphics{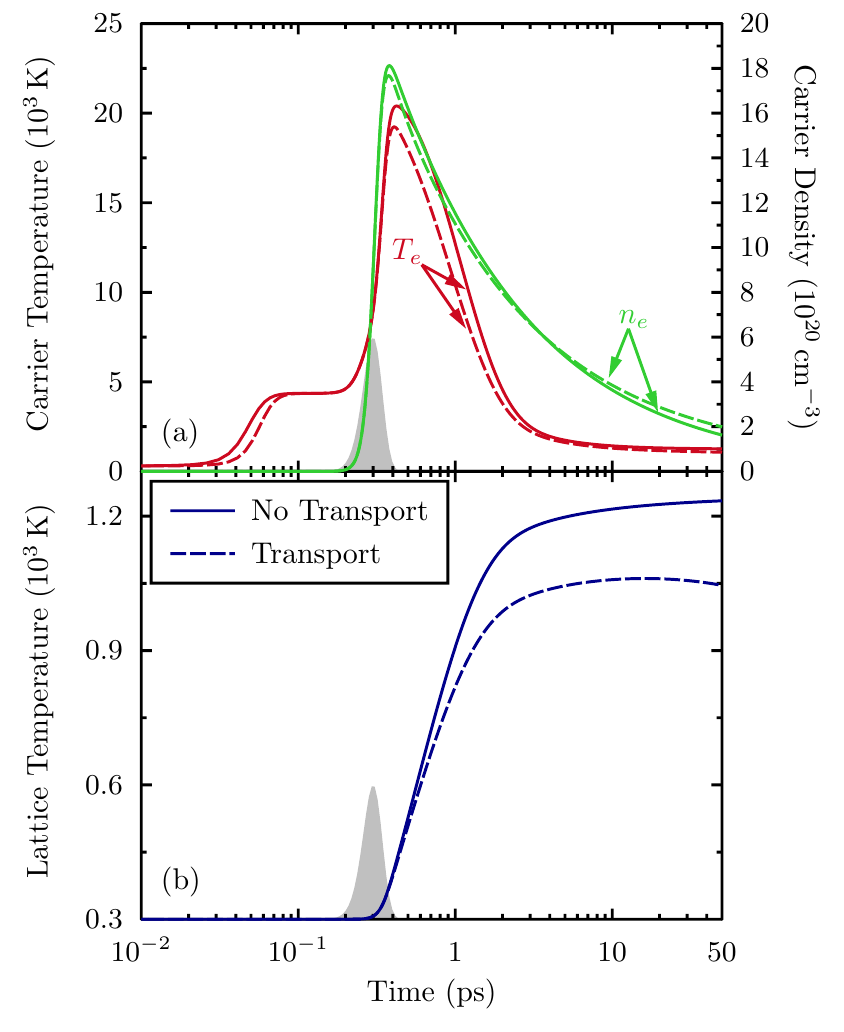}
				\caption{\label{fig:transport_surface_t-n-Te-Tp} (Color online) Carrier density and temperature (a) as well as lattice
				temperature (b) at the surface calculated either considering or neglecting transport.}
			\end{figure}

			Figure~\ref{fig:transport_surface_t-n-Te-Tp} depicts carrier and lattice temperature as well as carrier density at the
			surface calculated either considering or neglecting transport. The maxima of carrier and lattice temperature	
			are lower when transport is considered, because energy dissipates away from the surface. In addition, while	lattice
			temperature still increases even on long times scales of tens of picoseconds in the case without transport,
			it reaches a maximum and starts to decrease on longer timescales when transport is considered.

			For carrier density things are a bit different. 
			While the maximum carrier density shows the same behavior as the temperatures, after about \unit{5}{\pico\second}
			the density at the surface becomes \emph{higher} when transport is considered in comparison to the case without
			transport.
			This behavior of surface carrier density can be attributed to carrier confinement
			which has been investigated by Preston and van Driel\cite{vanDriel87,Preston84} for silicon irradiated with
			picosecond and nanosecond pulses.

			Let us now briefly have a look at the influence of transport effects on the estimated damage thresholds.
			
			\begin{figure}
				\includegraphics{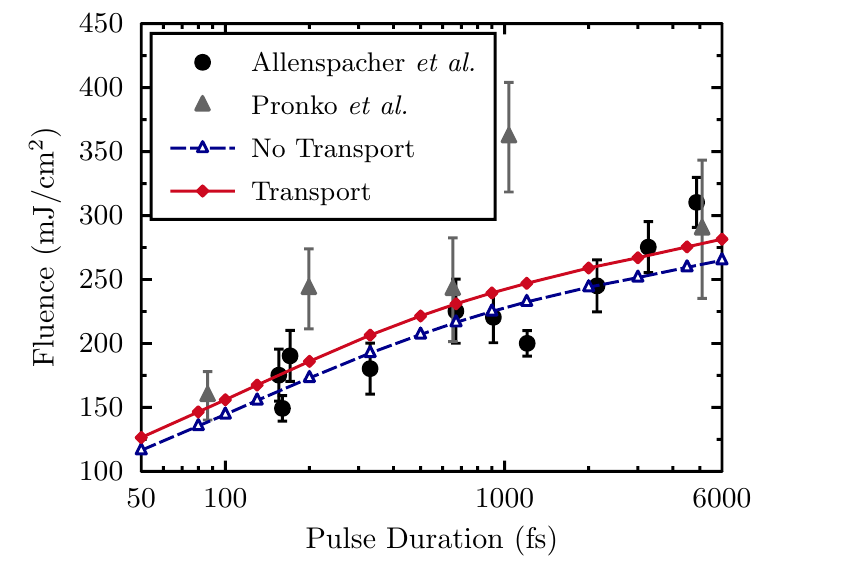}
				\caption{\label{fig:threshold_influence_of_transport} (Color online) Damage thresholds calculated either
				considering or neglecting transport in comparison with experimental data published by Allenspacher~\emph{et
				al.}\cite{Allenspacher03} and Pronko~\emph{et al.}\cite{Pronko98}}
			\end{figure}

			Figure~\ref{fig:threshold_influence_of_transport} depicts the damage thresholds calculated 
			using the full model (red diamonds, same curve appears in Fig.~\ref{fig:threshols_drude_vs_T},
			\ref{fig:threshold_influence_frequency} and \ref{fig:threshold_influence_degeneracy}) 
			and neglecting transport (blue triangles) in comparison with experimental data published by
			Allenspacher~\emph{et al.}\cite{Allenspacher03} and Pronko~\emph{et al.}\cite{Pronko98}

			The damage thresholds estimated neglecting transport are lower than those when considering transport but show the same
			overall behavior with increasing pulse duration and are still in good agreement with experimental data. The
			threshold being lower when neglecting transport can be explained by the fact that in this case no heat is carried
			away from the surface.

			Thus, we conclude that as long as only the temperature evolution at the surface is of interest (e.g. for the
			estimation of damage thresholds) and no detailed investigation of temperature or density profiles is necessary, 
			it is sufficient to implement an nTTM neglecting transport, which considerably reduces computational efforts.

	\section{Summary and Conclusion}

		In this work, we extended the nTTM first presented by van Driel\cite{vanDriel87} to account for the changes in optical
		parameters, namely reflectivity and FCA coefficient, due to the highly transient free carrier density during the
		excitation with femtosecond laser pulses. 

		For the irradiation of silicon with a \unit{100}{\femto\second}-laser pulse at \unit{800}{\nano\meter}, we analyzed the
		influence of the transient optical properties. We conclude, that it is of utter importance
		to consider changes in reflectivity and FCA coefficient due to the changing carrier density. 
		This can not only be seen in the reflectivity, temperature and density evolution but is also clearly
		reflected in the calculated damage thresholds. Comparison with experimental data shows that the $T$-expression 
		often used with the nTTM in earlier	works strongly overestimates the threshold.
		Damage thresholds calculated with our improved approach using a Drude model considering both 
		electron-hole
		and carrier-phonon collisions are, however, in 
		very good agreement with experimental data even over a wide range of pulse durations.
		Note that we did not fit any parameter of our model to reproduce experimental damage thresholds.

		Moreover, we found that the choice of carrier collision frequency used in the Drude model 
		strongly influences the lattice temperature and, consequently, the calculated damage threshold.
		Thus, we conclude that it does not suffice to assume a constant carrier collision frequency
		at least when treating a wide range of pulse durations. 
		In this case, a transient carrier collision frequency
		considering both, 
		electron-hole 
		collisions and carrier-phonon collisions, should be applied to best reproduce experimental data.
		
		We analyzed the influence of the distribution function assumed for the carriers and found that while both electron and
		hole system become degenerate when excited with the laser pulse, it suffices to assume a Maxwell-Boltzmann distribution
		for the carriers when estimating damage thresholds.
		
		Furthermore, we investigated transport of carriers and energy and found that the influence of the transient band 
		gap during the excitation can lead to carrier confinement in a region below the incident surface. 
		Moreover, we found that damage thresholds calculated neglecting transport are a bit lower than those when considering
		transport but show the same overall behavior with increasing pulse duration and are still in good agreement 
		with experimental data.

		We therefore conclude, that it suffices to use an nTTM neglecting transport and degeneracy effects.
		This immensely reduces computational effort in comparison to a full calculation.

		Finally, we conclude that our approach yields highly satisfying results in agreement with experimental data once 
		a Drude model considering a transient carrier collision frequency is implemented. The improved nTTM considering 
		transient optical parameters therefore is a powerful tool to describe laser-excited semiconductors.
		
	\begin{acknowledgments}
		The authors thank K. Sokolowski-Tinten and V. P. Lipp for helpful suggestions. Financial support by 
		the Deutsche Forschungsgemeinschaft through the Emmy Noether (grant no. RE 1141/11) and the Heisenberg program (grant
		no. RE 1141/15) is gratefully acknowledged.
	\end{acknowledgments}

%
\end{document}